\documentclass[10pt,letterpaper]{article}
\usepackage[top=0.85in,footskip=0.75in]{geometry}

\usepackage{changepage}

\usepackage[utf8x]{inputenc}

\usepackage{textcomp,marvosym}

\usepackage{fixltx2e}

\usepackage{amsmath,amssymb}

\usepackage{cite}

\usepackage{nameref,hyperref}

\usepackage[right]{lineno}

\usepackage{color}

\usepackage{tikz}
\usetikzlibrary{arrows}

\usetikzlibrary{decorations.markings}
\usetikzlibrary{decorations.pathmorphing}
\tikzstyle arrowstyle=[scale=1]
\tikzstyle directed=[{postaction={decorate,decoration={markings,
    mark=at position .65 with {\arrow[arrowstyle]{stealth}}}}, ultra thick}]
\tikzstyle directed_dotted=[{postaction={decorate,decoration={markings,
    mark=at position .65 with {\arrow[arrowstyle]{stealth}}}}, ultra thick,dotted}]
\tikzstyle synapse=[{decorate,decoration={snake}, ultra thick}]
\tikzstyle gray=[{circle,fill=gray,minimum size=10pt}]
\tikzstyle black=[{circle,fill=black,minimum size=10pt}]
\tikzstyle white=[{circle,draw,fill=white,minimum size=10pt}]
\tikzstyle math=[{circle,fill=none}]

\usepackage{microtype}
\DisableLigatures[f]{encoding = *, family = * }

\raggedright
\usepackage[aboveskip=1pt,labelfont=bf,labelsep=period,justification=raggedright,singlelinecheck=off]{caption}

\makeatletter
\renewcommand{\@biblabel}[1]{\quad#1.}
\makeatother

\date{}

\makeatletter
\newcommand*{\centerfloat}{%
  \parindent \z@
  \leftskip \z@ \@plus 1fil \@minus \textwidth
  \rightskip\leftskip
  \parfillskip \z@skip}
\makeatother

\usepackage{mdframed}

\begin{document}
\vspace*{0.2in}

\begin{flushleft}
{\Large
\textbf\newline{From the statistics of connectivity to the statistics of spike times in neuronal networks}  
}
\newline 
Gabriel Koch Ocker\textsuperscript{1*}, Yu Hu\textsuperscript{2*}, Michael A. Buice\textsuperscript{1,3}, Brent Doiron\textsuperscript{4,5}, Kre\v{s}imir Josi\'c\textsuperscript{6,7,8}, Robert Rosenbaum\textsuperscript{9}, Eric Shea-Brown\textsuperscript{3,1,10} \\


\bigskip
\textbf{1} Allen Institute for Brain Science.
\textbf{2} Center for Brain Science, Harvard University.
\textbf{3} Department of Applied Mathematics, University of Washington.
\textbf{4} Department of Mathematics, University of Pittsburgh.
\textbf{5} Center for the Neural Basis of Cognition, Pittsburgh.
\textbf{6} Department of Mathematics, University of Houston.
\textbf{7} Department of Biology and Biochemistry, University of Houston.
\textbf{8} Department of BioSciences, Rice University.
\textbf{9} Department of Mathematics, University of Notre Dame.
\textbf{10} Department of Physiology and Biophysics, and University of Washington Institute for Neuroengineering.
\textbf{*} Equal contribution.


%
%






\end{flushleft}

\section*{Highlights}

\begin{itemize}
\item Remarkable new data on  connectivity and activity raise the promise and raise the bar for linking structure and dynamics in neural networks.
\item Recent theories aim at a statistical approach, in which the enormous complexity of wiring diagrams is reduced to key features of that  connectivity that drive coherent, network-wide activity.
\item We provide a unified view of three branches of this work, tied to a broadly useful ``neural response'' formula that explicitly relates connectivity to spike train statistics. 
\item This isolates a surprisingly systematic role for the local structure and spatial scale of connectivity in determining spike correlations, and shows how the coevolution of structured connectivity and spiking statistics through synaptic plasticity can be predicted self-consistently.
\end{itemize}

\section*{Abstract}

An essential step toward understanding neural circuits is linking their structure and their dynamics.  In general, this relationship can be almost arbitrarily complex. Recent theoretical work has, however, begun to identify some broad principles underlying collective spiking activity in neural circuits.  The first is that local features of network connectivity can be surprisingly effective in predicting global statistics of activity across a network. The second is that, for the important case of large networks with excitatory-inhibitory balance, correlated spiking persists or vanishes depending on the {\it spatial scales} of recurrent and feedforward connectivity.  We close by showing how these ideas, together with plasticity rules, can help to close the loop between network structure and activity statistics.

\section{Introduction}
Here, we focus on relating network connectivity to collective activity at the level of spike times, or {\it correlations} in the spike counts of cells on the timescale of typical synapses or membranes (See {\bf Box 1}).  Such correlations are known to have complex but potentially strong relations with coding in single neurons \cite{dettner_temporal_2016} and neural populations~\cite{hu_sign_2014,moreno-bote_information-limiting_2014,zylberberg_direction-selective_2016, franke_structures_2016},  
and can modulate the drive to a downstream population \cite{kumar_spiking_2010}. 
Moreover, such correlated activity can modulate the evolution of synaptic strengths through spike timing dependent plasticity (STDP)(~\cite{markram_history_2011, ocker_self-organization_2015, tannenbaum_shaping_2016}, but see~\cite{graupner_natural_2016}). 

Collective spiking arises from two mechanisms:  connections among neurons within a population, and external inputs or modulations affecting the entire population~\cite{cohen_measuring_2011, mcginley_waking_2015, doiron_mechanics_2016}.  Experiments suggest that both are important.  Patterns of correlations in cortical micro-circuits have been related to connection probabilities and strengths~\cite{cossell_functional_2015}.  At the same time, latent variable models of dynamics applied to cortical data have revealed a strong impact of global inputs to the population~\cite{ecker_state_2014, rosenbaum_spatial_2017, goris_partitioning_2014, ecker_structure_2016}.  



At first, the path to understanding these mechanisms seems extremely complicated. Electron microscopy (EM) and allied reconstruction methods promise connectomes among thousands of nearby cells, tabulating an enormous amount of data ~\cite{lee_anatomy_2016, kasthuri_saturated_2015, bock_network_2011, kleinfeld_large-scale_2011, briggman_wiring_2011, helmstaedter_connectomic_2013, mishchenko_ultrastructural_2010}. This begs the question of what \emph{statistics} of connectivity matter most -- and least -- in driving the important activity patterns of neural populations.   The answer would give us a set of meaningful ``features'' of a connectome that link to basic statistical features of the dynamics that such a network produces.  Our aim here is to highlight recent theoretical advances toward this goal.

\section{Mechanisms and definitions:  sources and descriptions of (co)variability in spike trains}

\begin{figure}[h!]
\begin{mdframed}
\center{ \includegraphics[width=4 in]{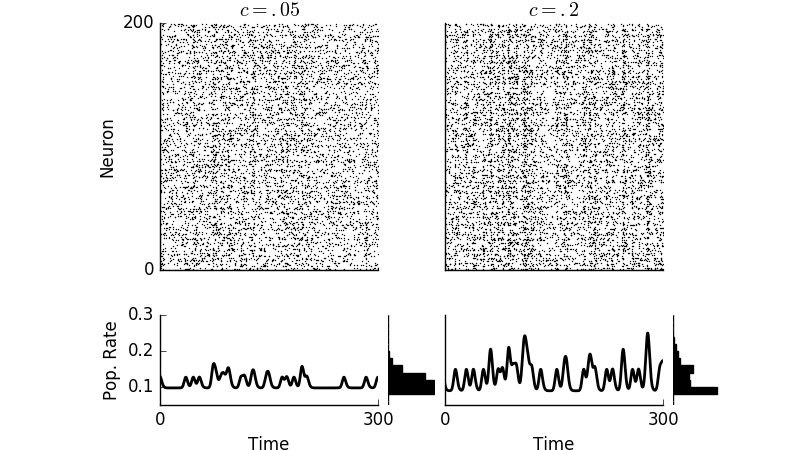}}
\caption*{ \small
{\bf Box 1.  Spike train statistics.} The spike train of neuron $i$ is defined as a sum of delta functions,
$y_i(t) = \sum_\mathrm{k} \delta \left(t - t_i^k\right).$ Spike train statistics can be obtained from samples of the spike trains of each neuron in a population. A joint moment density of $n$ spike trains is defined as a trial-average of products of those spike trains:
\begin{align} \label{E:moments}
\mathbf{m}_{i, j, \ldots, n}(t_i, \ldots, t_n) = \frac{1}{N} \sum_{\mathrm{trials}} \prod_{i=1}^n y_i(t_i),
\end{align}
where $N$ is the number of trials.  In practice, time is discretized into increments of size $\Delta t$, and spike trains are binned.  Equation~\eqref{E:moments} is recovered  from its discrete counterpart in the limit $\Delta t \rightarrow 0$. 
The first spike train moment is the instantaneous firing rate of a neuron $i$. The second spike train moment, $\mathbf{m}_{ij}(t_i, t_j),$ is the correlation function of the two spike trains; if $i=j$ it is an autocorrelation, otherwise a cross-correlation. It is frequently assumed that the spike trains are stationary, so that their statistics do not depend on time. We can then replace averages over trials with averages over time. Additionally, the correlation in this case only depends on the {\it time lag} in between spikes. This yields the spike train cross-correlation as,
\begin{align}
\mathbf{m}_{i, j, \ldots, n}(s_j, \ldots, s_n) = \frac{1}{T}\int_0^T dt_i \; \prod_{j=i+1}^n y_j(t_i + s_j),
\end{align}
where $s_j = t_j - t_i$  for $j=i+1, \ldots, n$ and $s_i$=0. The correlation function measures the frequency of spike pairs. Two uncorrelated Poisson processes with rates $r_i$ and $r_j$ have $\mathbf{m}_{ij}(s) = r_i r_j$, independent of the time lag $s$. The statistics of any linear functional of the spike trains (such as output spike counts, or synaptic outputs or inputs) can be derived from these spike train statistics \cite{rosenbaum_spatial_2017, ocker_linking_2016}.

Moments mix interactions of different orders. To account for lower-order contributions, we can define \emph{cumulants} of the spike trains. The first cumulant and the first moment both equal the instantaneous firing rate. The second cumulant is the covariance function of the spike train: $\mathbf{C}_{ij}(s) = \mathbf{m}_{ij}(s) - r_ir_j$. The third spike train cumulant similarly measures the frequency of triplets of spikes, above what could be expected by composing those triplets of individual spikes and pairwise covariances. Higher order cumulants have similar interpretations~\cite{novak_three_2012}.

}
\end{mdframed}
\end{figure}

Neurons often appear to admit spikes stochastically. Such variability can be due to noise from, e.g., synaptic release~\cite{faisal_noise_2008}, and can be internally generated via a chaotic ``balanced'' state ~\cite{van_vreeswijk_chaos_1996,van_vreeswijk_chaotic_1998,renart_asynchronous_2010}.
As a consequence, the structure of spike trains is best described statistically. The most commonly used statistics are the instantaneous firing rate of each neuron, the autocorrelation function of the spike train (the probability of observing pairs of spikes in a given cell separated by a time lag $s$), 
and the cross-correlation function (likewise, for spikes generated by two different cells).  As shown in {\bf Box 1}, even weak correlations yield coherent, population-wide fluctuations in spiking activity that can have a significant impact on cells downstream \cite{kumar_spiking_2010}.
Similarly, higher-order correlations are 
related to the probability of observing triplets, quadruplets or more spikes in a group of neurons, separated by a given collection of time lags. All these quantities correspond to {\it moment densities} of the spiking processes ({\bf Box 1}).

\section{Spike train covariability from recurrent connectivity and external input}
\label{sec:linearResponse}
In recent years, neuroscientists have advanced a very general framework for predicting how spike train correlations (more specifically, cumulants; {\bf Box 1}) depend on the structure of recurrent connectivity and external input.   
This framework is based on {\it linearizing the response} of a neuron around a baseline state of irregular firing. For simplicity we present the result as a  matrix of spike train \emph{auto-} and \emph{cross-spectra}, $\mathbf{C}(\omega)$; this is the matrix of the Fourier transforms of the familiar auto- and cross-covariance functions.  Under the linear response approximation, this is
\begin{align} \label{eq:TreeCov}
\mathbf{C}(\omega) 
= \underbrace{\mathbf{\Delta}(\omega) \mathbf{C}^0(\omega)\mathbf{\Delta}^*(\omega)}_\mathrm{Internally \; generated} + \underbrace{\mathbf{\Delta}(\omega)\left(\mathbf{A}(\omega)\mathbf{C}^\mathrm{ext}(\omega)\mathbf{A}^*(\omega) \right)\mathbf{\Delta}^*(\omega)}_\mathrm{Externally \; applied}.
\end{align}
%
The auto- (cross-) spectra correspond to diagonal (off-diagonal) terms of $\mathbf{C}(\omega)$. In particular, evaluated at $\omega = 0$, these terms give the variance and co-variance of spike counts over any time window large enough to contain the underlying auto and cross-correlation functions~\cite{rocha07,bair01}. The matrix $\mathbf{C}^0(\omega)$ is diagonal, and $\mathbf{C}^0_{kk}(\omega)$ is the power spectrum of neuron $k$'s spike train in the baseline state, without taking into account the effect of the recurrence on its spiking variance. $\mathbf{\Delta}^*$ denotes the conjugate transpose of $\mathbf{\Delta}$.

Here, the $ij^{th}$ entry of the matrix $\mathbf{\Delta}(\omega)$ is called a \emph{propagator}, and  reflects how a spike in neuron $j$ propagates through the network to affect activity in neuron $i$. When the activity can be linearized around a stable ``mean-field'' state, the matrix of propagators obeys:
\begin{align} \label{eq:propagator}
\mathbf{\Delta}(\omega) = \big(\mathbf{I} -\mathbf{K}(\omega) \big)^{-1},
\end{align}
where $\mathbf{K}(\omega) = \mathbf{A}(\omega)\mathbf{W}$ is the interaction matrix, which, importantly, encodes weights of synaptic connections $\mathbf{W}_{ij} $ between neurons $j$ and $i$. In general, the connection matrix can have a time dependence corresponding to the filtering and delay of synaptic interactions, and be written as $\mathbf{W}(\omega)$, but we suppress this $\omega$-dependence for ease of notation. 
$\mathbf{A}$ is a diagonal matrix; $\mathbf{A}_{ii}(\omega)$ is the linear response (Fourier transform of its impulse response, or PSTH~\cite{gabbiani_mathematics_2010}) of neuron $i$ to a perturbation in its synaptic input. 

The first term of Eq.~\eqref{eq:TreeCov} thus measures how the variability in the spiking of each neuron propagates through the network to give rise to co-fluctuations in pairs of neurons downstream.  These correlations thus generated from within the population; the second term in Eq. \eqref{eq:TreeCov} 
captures correlations generated by external inputs. 

The external inputs are described by their power spectrum, $\mathbf{C}^\mathrm{ext}_{ij}(\omega)$.  A global modulation in the activity of many neurons due to shifts in attention, vigilance state, and/or motor activity, would result in low-rank matrix $\mathbf{C}^\mathrm{ext}$. In this case the second, external term of Eq. \eqref{eq:TreeCov} will itself be low rank, since the rank of a matrix product $AB$ is bounded above by the ranks of $A$ and $B$.  Experimentally obtained  spike covariance matrices can be decomposed into a low-rank and ``residual'' terms~\cite{yatsenko_improved_2015,ecker14} that correspond to the two terms in the matrix decomposition Eq.~\eqref{eq:TreeCov}.

In the simplest case of an uncoupled pair of neurons $i$ and $j$ receiving common inputs, Eq.~\eqref{eq:TreeCov} reduces to $
\mathbf{C}_{ij}(\omega) = \mathbf{A}_i(\omega) \mathbf{C}^\mathrm{ext}_{ij}(\omega) \mathbf{A}_j(-\omega)$ \cite{de_la_rocha_correlation_2007,shea-brown_correlation_2008}. The covariance of the two spike trains is thus given by the input covariance, multiplied by the gain with which each neuron transfers those common inputs to its output. 

Eq.~\eqref{eq:propagator} can be expanded in powers of the interaction matrix $\mathbf{K}(\omega)$ as
\begin{align}
\mathbf{\Delta}(\omega) = \sum_{m=0}^\infty \mathbf{K}^m(\omega).
\end{align}
This expansion has a simple interpretation: $\mathbf{K}^m_{ij}(\omega)$ represents paths from a neuron $i$ to neuron $j$ that are exactly $m$ synapses long (with synaptic weights $\mathbf{W}$ weighted by the postsynaptic response gain $\mathbf{A}(\omega)$)~\cite{pernice_how_2011,trousdale_impact_2012}. Using this expansion in Eq.~\eqref{eq:TreeCov}, without external input, also provides an intuitive description of the spike train cross-spectra in terms of paths through the network,
\begin{align} \label{eq:expandK}
\mathbf{C}_{ij}(\omega) \approx \sum_{k} \sum_{m=0}^X\sum_{n=0}^Y \underbrace{\Big(\mathbf{K}^m(\omega)\Big)_{ik} \Big(\mathbf{K}^n (-\omega) \Big)_{jk}}_{\text{path terms}} \mathbf{C}^0_{kk}(\omega).
\end{align}
%
This expression explicitly captures contribution to the cross-spectrum, $\mathbf{C}_{ij}(\omega),$ of all paths of up to $X$ synapses ending at neuron $i$, and all paths of up to $Y$ synapses ending at neuron $j$. The index $k$ runs over all neurons. A first step toward the path expansion was taken by Ostojic, Brunel \& Hakim, who explored the first-order truncation of Eq.~\eqref{eq:TreeCov} for two cells, thus capturing contributions from direct connections and common inputs ~\cite{ostojic_how_2009}.

The formulation of Eq.~\eqref{eq:TreeCov} has a rich history.  Sejnowski used a similar expression in describing collective fluctuations in firing rate networks \cite{sejnowski_stochastic_1976}.  Linder, Doiron, Longtin and colleagues derived this expression in the case of stochastically driven integrate and fire neurons \cite{doiron_oscillatory_2004,lindner_theory_2005}, an approach generalized by  Trousdale et al.~\cite{trousdale_impact_2012}.  Hawkes derived an equivalent expression for the case of linearly interacting point processes (now called a multivariate Hawkes process), as pointed out by Pernice et al., who applied and directly related them to neural models \cite{hawkes_spectra_1971,pernice_how_2011,pernice_recurrent_2012}.  The correspondence of the rate dynamics of the Hawkes model, networks of integrate-and-fire neurons, and binary neuron models, was discussed in detail in~\cite{grytskyy_unified_2013} and the direct approximation of integrate-and-fire neurons by linear-nonlinear-Poisson models in~\cite{ostojic_spiking_2011}.

Moreover, Buice and colleagues \cite{buice_systematic_2010} developed a field theoretical method that encompasses the above approach and extends the formulation to correlations of arbitrary order.  Importantly, this also allows for nonlinear interactions.  An expansion can be derived via this method that describes the coupling of higher order correlations to lower moments:  in particular, pairwise correlations can impact the activity predicted from mean field theory~\cite{ocker_linking_2016}.  The field theoretic approach has also been applied to models of coupled oscillators related to neural networks, specifically the Kuramoto model \cite{hildebrand_kinetic_2007, buice_correlations_2007} and networks of ``theta" neurons \cite{buice_dynamic_2013}.  Similarly, Rangan developed a motif expansion of the operator governing the stationary dynamics of an integrate-and-fire network \cite{rangan09a}.

\section{Network motifs shape collective spiking across populations}

The relationship of spike train cross-spectra to pathways through the network provides a powerful tool for understanding how network connectivity shapes the power spectrum of the population-wide network activity ({\bf Box 1}).
The population power spectrum is given by the average over the cross-spectral matrix: $C(\omega) = \left \langle \mathbf{C}_{ij}(\omega) \right \rangle_{i,j}$ (the angle brackets denote averaging over pairs of neurons within a given network). 
Therefore $C(\omega)$ is the average of the left-hand side of Eq.~\eqref{eq:expandK}. We show that the right-hand side of Eq.~\eqref{eq:expandK} can in turn be linked to the \emph{motif moments} which describe the mean strength of different weighted microcircuits in the network. 

\begin{figure}
\centering \includegraphics[width=6 in]{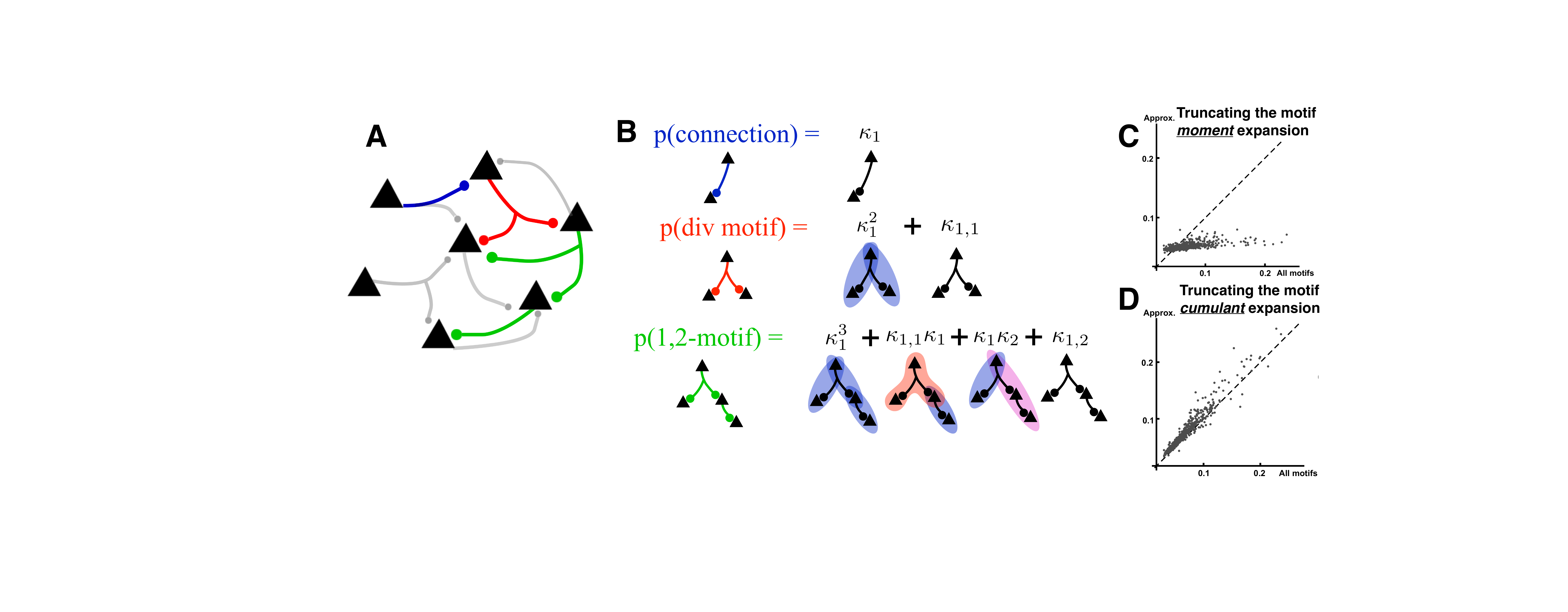}
\caption{(A) Various motifs are identified throughout the neural network. Their frequency can be measured by counting their occurrence. (B) The probability of a motif (motif moments $\mu_{m,n}$, see text) can be decomposed into cumulants of smaller motifs. (C) Comparing the average correlation between excitatory neurons calculated using motif statistics of all orders ($X,Y\rightarrow \infty$ in Eq.~\eqref{eq:expandK}) and approximations using up to second order ($X+Y\leq 2$) for 512 networks (different dots) with various motif structures. Each network is composed of 80 excitatory and 20 inhibitory neurons. The deviation from the dashed line $y=x$ shows that motif moments beyond second order are needed to accurately describe correlations. (D) Same as (C) for a motif cumulant expansion (Eq.~(44) in \cite{hu_motif_2013}, a generalization of the motif cumulant theory for networks with multiple neuron populations) truncated after second order motif terms; both panels adapted from~\cite{hu_motif_2013}.} \label{fig:motif}
\end{figure}

Assuming cellular response properties are homogeneous, the interaction matrix can be written as $\mathbf{K}(\omega) = A(\omega) \mathbf{W}$, where the scalar kernel $A(\omega)$ is the same response kernel for all cells. 
If the baseline auto-correlations, hence $\mathbf{C}^0_k$, are also equal across the network, the ``path terms'' appearing in Eq.~\eqref{eq:expandK} are directly proportional to the {\it motif moments} of the connectivity matrix $\mathbf{W},$ defined as:
\begin{align}
\mu_{m,n} = \left \langle \mathbf{W}^m \left(\mathbf{W}^T\right)^n \right \rangle_{i,j} / N^{m+n+1}.
\label{eq:def_motif_moment}
\end{align}
This measures the average strength of a $(m,n)$--motif composed of two paths of synapses emanating from a neuron $k$ with one path of $m$ synapses ending at neuron $i$, the other path of $n$ synapses ending at neuron $j$. 
Examples of a $(1,1)$-, and $(1,2)$-motif are shown in Fig.~\ref{fig:motif}A. For networks where $\mathbf{W}_{ij}= 0$ or $1$, $\mu_{m,n}$ is also the frequency of observing a motif in the network. Eq.~\eqref{eq:expandK} thus provides a way to approximate how average correlations depend on the frequency of motifs in the network.

An accurate approximation of the population power spectrum typically requires keeping many terms in the series (Fig.~\ref{fig:motif}C).  This is a challenge for applications to real neural networks, where the statistics of motifs involving many neurons are much harder to determine. Importantly, however, the contribution of these  higher order motifs can often be decomposed into contributions of smaller, component motifs by introducing~\emph{motif cumulants} (Fig.~\ref{fig:motif} B) \cite{hu_motif_2013,hu_local_2014}. This approach allows us to remove  redundancies in motif statistics, and isolate the impact solely due to higher order motif structure.
This motif cumulant expansion allows Eq.~\eqref{eq:expandK} to be re-arranged in order to only truncate higher-order motif cumulants, rather than moments, providing a much-improved estimation of the spike train covariances (Fig. \ref{fig:motif}C vs D, \cite{hu_motif_2013}) that is still based only on very small motifs.

In sum, population-averaged correlation can be efficiently linked to \emph{local connectivity structures} described by motif cumulants. Significant motif cumulant structure exists in local networks of both cortex \cite{song_highly_2005,perin_synaptic_2011} and area CA3 of the hippocampus, where they play a crucial role in pattern completion \cite{guzman_synaptic_2016}. Moreover, as we see in Sec.~\ref{sec:plasticity}, correlations in turn can shape a network's motif structure through synaptic plasticity.

\paragraph*{Higher-order correlations and network structure}
While we have discussed how network structure gives rise to correlations in pairs of spike trains, joint activity in larger groups of neurons, described by higher-order correlations, can significantly affect population activity \cite{
ohiorhenuan_sparse_2010, shimazaki_state-space_2012, 
tkacik_searching_2014}. Analogous results to Eq.~\eqref{eq:TreeCov} exist for higher-order correlations in stochastically spiking models \cite{buice_systematic_2010, buice_beyond_2013, jovanovic_cumulants_2015, ocker_linking_2016}. Network structure has been linked to the strength of third-order correlations in networks with narrow degree distributions \cite{jovanovic_interplay_2016}, and the allied motif cumulant theory developed~\cite{HuCosyne15}, advancing the aim of understanding how local connectivity structure impacts higher-order correlations across networks. Finally, a range of work has characterized the statistics of avalanches of neuronal activity: population bursts with power law size distributions \cite{plenz_organizing_2007}, which potentially suggest a network operating near an instability \cite{BuiceC07}. 

\paragraph*{Motifs and stability}
The results discussed so far rely on an expansion of activity around a baseline state where neurons fire asynchronously. How such states arise is a question addressed in part by the mean-field theory of spiking networks \cite{ginzburg_theory_1994, van_vreeswijk_chaos_1996, brunel_dynamics_2000, renart_asynchronous_2010, tetzlaff_decorrelation_2012, helias_correlation_2014}. The existence of a stable stationary state depends on the structure of connectivity between neurons. In particular, when connectivity is strong and neurons have heterogeneous in-degrees, the existence of a stable mean-field solution can be lost \cite{pyle_highly_2016, landau_impact_2016}. One way to rescue a stable activity regime is to introduce correlations between neurons' in and out-degrees \cite{pyle_highly_2016}; these correspond to chain motifs ($\kappa_{2}$ in Fig. 1B). Therefore the motifs that control correlated variability also affect the stability of asynchronous balanced states.   

Motif structure also affects oscillatory population activity. Roxin showed that in a rate model generating oscillations of the population activity, the variance of in-degrees (related to the strength of convergent motifs) controls the onset of oscillations \cite{roxin_role_2011}. Zhao et al. took a complementary approach of examining the stability of completely asynchronous and completely synchronous states, showing that two-synapse chains and convergent pairs of inputs regulate the stability of completely synchronous activity \cite{zhao_synchronization_2011}. 



\section{Spatial scale of connectivity and inputs determines correlations in large networks}




Cortical neurons receive strong external and recurrent excitatory projections that would, if left unchecked, drive neuronal activity to saturated levels.  Fortunately, strong recurrent inhibition {\it balances} excitation, acting to stabilize cortical activity and allow moderate firing.  These large and balanced inhibitory and excitatory inputs are a major source of synaptic fluctuations, ultimately generating output spiking activity with Poisson-like variability \cite{doiron_balanced_2014,deneve2016}. 

A central feature of balanced networks is that they produce {\it asynchronous}, uncorrelated spiking activity (in the limit of large networks).   Original treatments of balanced networks by van Vreeswijk and Sompolinsky \cite{van_vreeswijk_chaos_1996,van_vreeswijk_chaotic_1998} and Amit and Brunel \cite{amit97} explained asynchronous activity by assuming sparse wiring within the network, so that shared inputs between neurons were negligible.  While the connection probability between excitatory neurons is small~\cite{oswald09,Holmgren03,lefort2009}, there is abundant evidence that connections between excitatory and inhibitory neurons can be quite dense \cite{fino_dense_2011,oswald09}. Renart, de la Rocha, Harris and colleagues showed that homogeneous balanced networks admit an asynchronous solution despite dense wiring \cite{renart10} (for large networks). This result suggests a much deeper relationship between balance and asynchronous activity than previously realized.  Building on this work, Rosenbaum, Doiron and colleagues extended the theory of balanced networks to include spatially dependent connectivity \cite{rosenbaum_balanced_2014,rosenbaum2017}.  We review below how the spatial spread of connectivity provides new routes to correlated activity in balanced networks.
       
\begin{figure}[h!] 
\vspace{-1mm}
\centering
\includegraphics[width=6.5in]{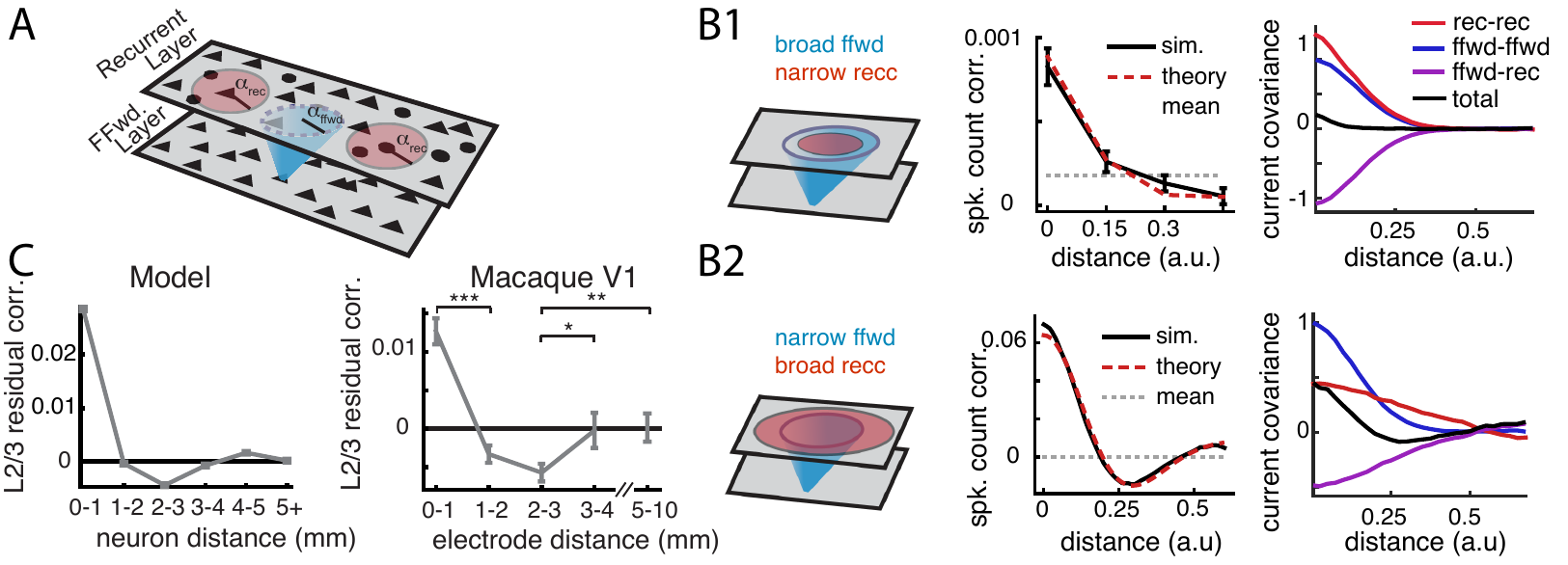}
\caption{Correlated activity in balanced networks with spatially dependent connections. (A) Schematic of the two layer network.  The blue/red zones denotes the spatial scale of feedforward/recurrent connectivity. (B1) Broad feedforward and narrow recurrent connectivity (left) produce an asynchronous state (middle).  The asynchrony requires a cancellation of $C_{RR}(d)$ and $C_{FF}(d)$ by $C_{RF}(d)$ at all distances (right). (B2) Narrow feedforward and broad recurrent connectivity (left) produce spatially structured correlations (middle) because $C_{RF}(d)$ does not cancel $C_{RR}(d)$ and $C_{FF}(d)$ (right). (C) When a one dimensional latent variable is extracted and removed from the primate V1 array data, the model predictions (left) are validated (right).  Panels are from \cite{rosenbaum2017}.}
\label{Fig_balanced_space}
\end{figure}

Consider a two-layer network, with the second layer receiving both feedforward ($F$) and recurrent ($R$) inputs (Fig. \ref{Fig_balanced_space}A).  For simplicity we assume that the feedforward and recurrent projections have Gaussian profiles with widths $\sigma_{\textrm{F}}$ and $\sigma_{\textrm{R}}$, respectively.  Each neuron receives the combined input $I=F+R$, decomposing the input covariance $C_{II}(d)$ to a representative pair of layer two neurons separated by a distance $d$ as:  
\begin{equation}\label{E:balanced_asych}
\begin{aligned} 
C_{II}(d)=C_{FF}(d)+C_{RR}(d)+2C_{RF}(d).
\end{aligned}
\end{equation} 
Here $C_{FF}$ and $C_{RR}$ are the direct covariance contributions from feedforward and recurrent pathways, respectively, while $C_{RF}$ is the indirect contribution to covariance from the recurrent pathway tracking the feedforward pathway.   

If the network coupling is dense and layer one neurons are uncorrelated with one another then $C_{FF}$, $C_{RR}$ and $C_{RF}$ are all $\mathcal{O}(1)$.  This means that feedforward and recurrent projections are potential sources of correlations within the network.  The asynchronous state requires that $C_{II}$ $\sim$ $\mathcal{O}(1/N)$.  This can only be true if the feedforward and recurrent correlations are {\it balanced} so that the recurrent pathways tracks and cancels the correlations due to the feedfoward pathway.  If we take $N \to \infty$ then in the asynchronous state, $C(d) \to 0$ implies: 
\begin{equation}\label{E:C_RF}
\begin{aligned} 
C_{RF}(d) = -\frac{1}{2} \left (C_{FF}(d)+C_{RR}(d) \right ).
\end{aligned}
\end{equation}
This must be true for {\it every distance $d$}, and from Eq. \eqref{E:balanced_asych} we derive \cite{rosenbaum2017} that the various spatial scales must satisfy: 
\begin{equation}\label{E:balanced_sigmas}
\begin{aligned} 
\sigma^2_{\textrm{F}}=\sigma^2_{\textrm{R}}+\sigma^2_{\textrm{\textrm{rate}}}.
\end{aligned}
\end{equation}
Here $\sigma^2_{\textrm{\textrm{rate}}}$ is the spatial scale of correlated firing within the network.  The intuition here is that for cancellation at every $d$ the spatial scale of feedforward and recurrent correlations must match one another.  While the spatial scale of $C_{FF}(d)$ is determined only by $\sigma_{F}$, the scale of recurrent correlations is calculated from the correlated spiking activity convolved with the recurrent coupling (hence the sum $\sigma^2_{\textrm{R}}+\sigma^2_{\textrm{\textrm{rate}}}$).    While $\sigma_{F}$ and $\sigma_{R}$ are architectural parameters of the circuit (and hence fixed), $\sigma_{\textrm{rate}}$ is a model output that must be determined. For any solution to make sense we require that $\sigma_{\textrm{rate}} >0$.  This gives a compact asynchrony condition: $\sigma_{F} > \sigma_{R}$.  In other words for feedforward and recurrent correlations to cancel, the spatial spread of feedforward projections must be larger than the spatial spread of recurrent projections.  

To illustrate how the spatial scales of connectivity control the asynchronous solution, we analyze the activity of a balanced spiking network when $\sigma_{F} > \sigma_{R}$ is satisfied (Fig. \ref{Fig_balanced_space}B1, left).  As expected, the spiking activity is roughly asynchronous with spike count correlations  near zero  (Fig. \ref{Fig_balanced_space}B1, middle).  When we examine the contributions of the feedforward and recurrent pathways, we see that the relation in Eq. \eqref{E:C_RF} is satisfied (Fig. \ref{Fig_balanced_space}B1, right).  We contrast this to the case when $\sigma_{F} < \sigma_{R}$, violating the  asynchrony condition (Fig. \ref{Fig_balanced_space}B2, left).  Here, a clear signature of correlations is found: neuron that are nearby one another are positively correlated while more distant neuron pairs are negatively correlated.  Indeed, the cancellation condition Eq. \eqref{E:C_RF} is violated at almost all $d$ (Fig. \ref{Fig_balanced_space}B2, right).      

Layer 2/3 of macaque visual area V1 is expected to have $\sigma_{F} < \sigma_{R}$, with long range projections within 2/3 being broader than L4 projections to L2/3 \cite{bosking1997,lund_anatomical_2003}.  Smith and Kohn collected population activity over large distances in layer 2/3 of macaque V1 \cite{smith08}.  When a one dimensional source of  correlations is removed from the data then the model prediction is supported (Fig. \ref{Fig_balanced_space}C).  

While these arguments give conditions for when the asynchronous solution will exist, it cannot give an estimate of correlated activity.  Rather, when either $N < \infty$ or $\sigma_{F} < \sigma_{R}$, we require the linear response formulation of Eq.~\eqref{eq:TreeCov} to give a prediction of how network correlations scale with distance (red dashed in Fig. \ref{Fig_balanced_space}B1 and B2, middle).  It is interesting to note that a majority of the fluctuations are internally generated within the balanced circuit and subsequently have a rich spectrum of timescales.  For spikes  counted over long windows, the linear response formalism of Eq.~\eqref{eq:TreeCov} once again predicts the underlying correlations (``theory" curves).

\section{Joint activity drives plasticity of recurrent connectivity}
\label{sec:plasticity}
\begin{figure}[h!] 
\vspace{-1mm}
\centering
\includegraphics[width=6.5in]{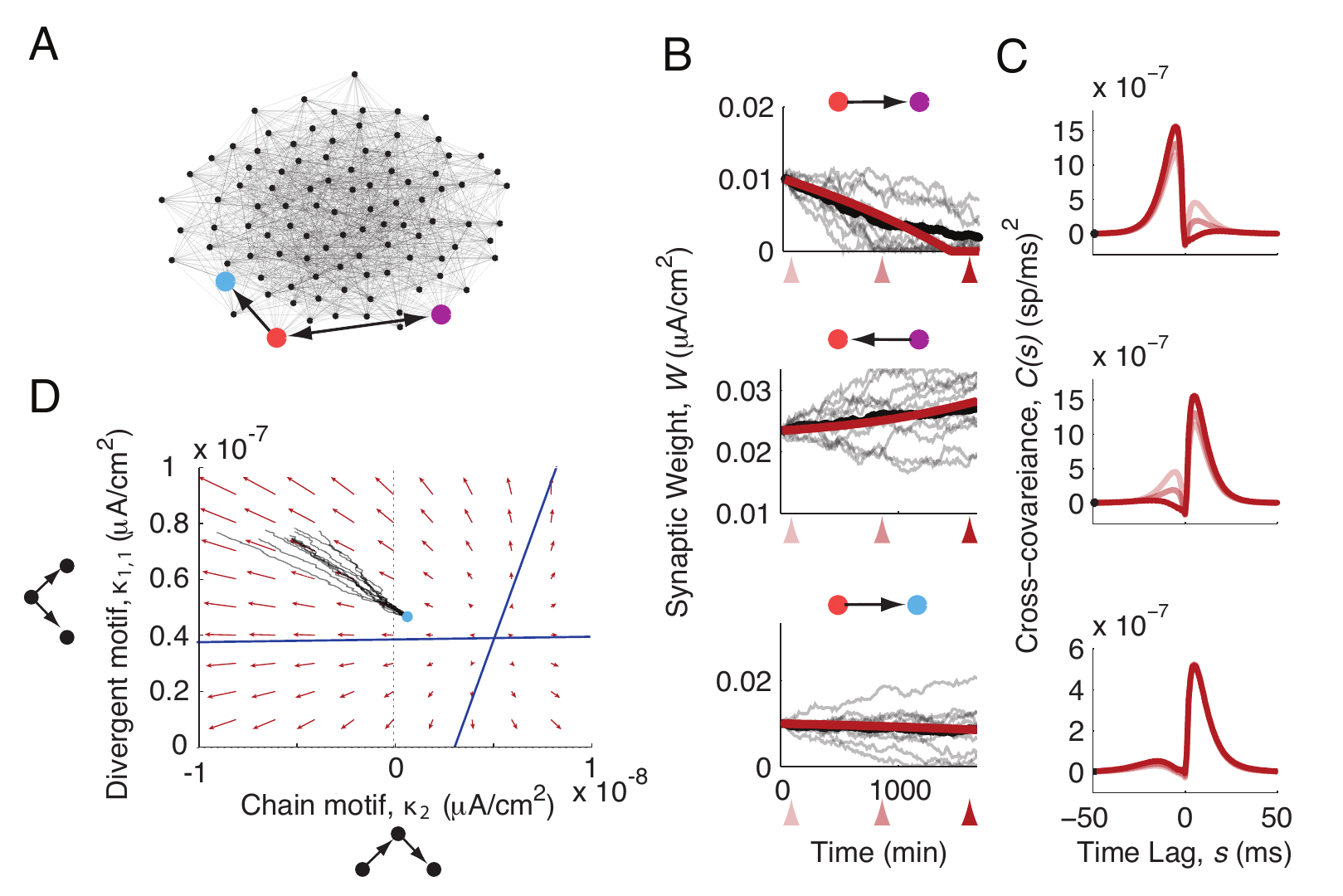}
\caption{Spike timing-dependent plasticity gives rise to joint evolution of synaptic weights, spike train covariances, and motif statistics. Panels from Ocker et al., 2015. A) Diagram of network structure. B) Evolution of the synapses highlighted in panel A. B) Evolution of the spike train covariances between the pre- and postsynaptic cells of each synapse, predicted using the first-order truncation of Eq. \eqref{eq:expandK}. Shading corresponds to the time points marked with arrows below the time axis of panel B. D) Projection of the joint dynamics of two-synapse motifs into the (divergent, chain) plane under a pair-based, additive Hebbian plasticity rule.}
\vspace{-4mm}
\label{fig:plasticity}
\end{figure}

The structure of neuronal networks is plastic, with synapses potentiating or depressing  in a way that depends on pre- and postsynaptic spiking activity~\cite{feldman_spike-timing_2012}. When synaptic plasticity is slow compared to the timescales of spiking dynamics, changes in  synaptic weights are linked to the {\it statistics} of the spike trains \cite{kempter_hebbian_1999}:  specifically,  the joint moment densities of the pre- and postsynaptic spike trains~\cite{gerstner_mathematical_2002} ({\bf Box 1}). For plasticity rules based on pairs of pre- and postsynaptic spikes, this results in a joint evolution of the weight matrix $\mathbf{W},$ the firing rates, $\vec{r}$, and the cross-covariances, $\mathbf{C}(s)$ (Fig. \ref{fig:plasticity}A-C). 

As we saw above, spike train cumulants depend on the network structure. In the presence of plasticity mechanisms, the structure of neuronal networks thus controls its own evolution -- both directly by generating correlations \cite{ocker_self-organization_2015,tannenbaum_shaping_2016} and indirectly, by filtering the correlations inherited from external sources \cite{gilson_emergence_2009-1, ocker_training_2016}. Recent work has leveraged this connection to determine how particular structural motifs shape spike train correlations to drive plasticity~\cite{tannenbaum_shaping_2016}. A further step is to close the loop on motifs, leveraging approximations of the true spike-train correlations in order to predict the plasticity dynamics of motif cumulants \cite{ocker_self-organization_2015, ocker_training_2016}:
\begin{align}
\frac{d}{dt} \vec{\kappa}  = F\left(\mathbf{m}\left(\vec{\kappa}, \mathbf{C}^\mathrm{ext}\right)\right),
\end{align}
where $\vec{\kappa}$ represents a chosen set of motif cumulants and the form of the function $F$ depends on the plasticity model used. Such analysis reveals that under an additive, pair-based plasticity rule where pre-post pairs cause potentiation and post-pre pairs cause depression, an unstructured weight matrix (with zero motif cumulants) is unstable: motifs will spontaneously potentiate or depress, creating structure in the synaptic weights (Fig. \ref{fig:plasticity}D),~\cite{ocker_self-organization_2015}. So far, such studies have focused on plasticity driven by spike pairs, relying on the linear response theories of \cite{Hawkes71a, pernice_how_2011, trousdale_impact_2012}. More biologically realistic plasticity models rely on multi-spike interactions and variables measuring postsynaptic voltage or calcium concentrations \cite{markram_history_2011}.
Theories describing higher-order spike-train and spike-voltage or spike-calcium correlations provide a new window through which to examine networks endowed with these richer plasticity rules.

\section{Conclusion}

The next few years could be pivotal for the study of how network structure drives neural dynamics.  Spectacular experimental methods are producing vast datasets that unite connectivity and activity data in new ways~\cite{ cossell_functional_2015, bock_network_2011, lee_anatomy_2016, briggman_wiring_2011}.  Does our field have a theory equal to the data?  We've reviewed how mathematical tools can separate the two main mechanisms giving rise to collective spiking activity -- recurrent connectivity and common  inputs -- and how this connects with decomposition methods in data analysis ~\cite{ecker_state_2014,goris_partitioning_2014,rosenbaum_spatial_2017,yatsenko_improved_2015}.  Moreover, we are beginning to understand how local and spatial structures scale up to global activity patterns, and how they drive network plasticity.  Daunting challenges remain,  from the role of myriad cell types to the impact of nonlinear dynamics.  As the bar ratchets up ever faster, the field will be watching to see what theories manage to clear it.



\section*{Acknowledgements }

We acknowledge the support of the NSF through grants DMS-1514743 and DMS-1056125 (ESB), DMS-1313225 and DMS-1517082 (B.D.), DMS-1517629 (KJ), DMS-1517828 (R.R.) and NSF/NIGMS-R01GM104974 (KJ), as well as the Simons Fellowships in Mathematics (ESB and KJ). GO, MB, and ESB wish to thank the Allen Institute founders, Paul G. Allen and Jody Allen, for their vision, encouragement and support.

\newpage

\section{Highlighted References}

\begin{itemize}

\item ** Pernice et al., PLOS Comp. Biol., 2011.  Calculates  the full matrix of pairwise correlation functions either in networks of  interacting Poisson (Hawkes process) neurons.  Shows how network-wide correlation is related to connectivity paths and motifs. 

\item *Trousdale et al., PLOS Comp. Biol, 2012. Calculates the linear response predictions for the full matrix of pairwise correlation functions, and their relationship to connectivity paths, in networks of integrate-and-fire neurons.  

\item ** Jovanovic \& Rotter, PLOS CB 2017. Building on their PRE 2015 paper, calculates the linear response predictions for the $n$-th order cumulant densities of multivariate Hawkes processes (i.e. linear-Poisson neural networks), and shows how these are related to connectivity in regular networks (those with narrow degree distributions).

\item ** Ocker et al, ArXiV 2016.  Extends existing theories for spike correlations to allow for nonlinear interactions among inputs and firing rates, deriving series of ``diagrams'' that show how connectivity influences activity statistics for quadratic and higher orders firing rate responses.

\item ** Hu et al., J Stat Mech 2013. Decomposes the effects of higher order motifs on spike correlations into smaller ones, and show how this {\it motif cumulant} approach enables one to predict global, population-wide correlations from the statistics of local, two-synapse motifs. 

\item * Hu et al., PRE 2014 .  Formulates network motif decompositions via a combinatorial relationship between moments and cumulants, extends the theory to multi-branch motifs related to higher order correlations, and to the impact of heterogeneous connectivity.

\item ** Renart, de la Rocha et al. Science 2010. Shows that for balanced networks with dense and strong connectivity, an asynchronous state emerges in the large-N limit in which excitatory-inhibitory interactions dynamically cancel excitatory-excitatory and inhibitory-inhibitory correlations. This leads to spike correlations that vanish on average. Also in the large N limit, this arises from population responses that instantaneously and linearly track external inputs.

\item ** Tetzlaff, Helias et al., PLoS CB 2012. A highly complementary study to the more well-known work of Renart et al. Using linear response theory for finite-size integrate-and-fire networks, Tetzlaff, Helias et al. expose  negative feedback loops in the dynamics of both purely inhibitory and excitatory-inhibitory networks, which give rise to the dynamical cancellation of correlations in finite-size networks.

\item * Helias et al., PLoS CB 2014. Building on the study of Renart, de la Rocha et al. and on their previous work, the authors disentangle correlation cancellation by excitatory-inhibitory interactions (reflected in the suppression of fluctuations in the population activity) from the tracking of external inputs.

\item ** Rosenbaum et al., Nature Neuroscience 2016. Building on their 2014 work in PRX, the authors extends the theory of correlations in balanced networks to systems with spatially dependent connectivity, showing that mismatches in the spatial scales of feedforward and recurrent inputs can give rise to stable average firing rates but  significant spike correlations. 

\item *Rangan, PRL 2009. Develops a diagrammatic expansion for the statistics of stationary integrate-and-fire networks with delta synapses in terms of subnetworks (i.e. motifs) via a functional representation of the full network dynamics. (See also the accompanying PRE article)

\item ** Zhao et al, Frontiers in Comp. Neuroscience, 2011.  A pioneering study of how network motifs impact network-wide synchrony, including simulations, analytical results, and methods of  generating useful networks to test theories linking structure to activity.

\item ** Gilson et al., Biol. Cyb. 2009iv. Building on the authors' previous work on STDP in networks of Poisson neurons (Gilson et al., Biol. Cyb. 2009i-iii), uses a combination of the Hawkes theory and a mean-field calculation of synaptic weights to show that externally-generated correlations can give rise to selective connectivity in recurrent networks of Poisson neurons through spike timing-dependent plasticity.

\item  ** Ocker et al., PLoS CB 2015. Uses the linear response theory for integrate-and-fire neurons to construct self-consistent predictions for internally-generated correlations and synaptic plasticity, and then derive a reduced dynamics for the spike timing-dependent plasticity of 2-synapse motif cumulants. This shows that an initially unstructured (Erd\H{o}s-R\`enyi) network connectivity is unstable under additive, Hebbian STDP unless all synapses potentiate or depress together.

\end{itemize}

\bibliography{sda_bib.bib,Brent.bib,cosyne.bib}



\end{document}